# High-throughput property-driven generative design of functional organic molecules

Julia Westermayr,[1,2] Joe Gilkes,[1,3] Rhyan Barrett,[1,2] Reinhard J. Maurer[1]

[1)] Department of Chemistry, University of Warwick, Gibbet Hill Rd, CV4 7AL, Coventry, UK,
[2)] Current address: Wilhelm-Ostwald-Institut für Physikalische und Theoretische Chemie, Universität Leipzig, Linnéstraße 2, D-04103 Leipzig, Germany
[3)] HetSys Centre for Doctoral Training, University of Warwick, Gibbet Hill Rd, CV4 7AL, Coventry, UK
julia.westermayr@uni-leipzig.de, r.maurer@warwick.ac.uk

The design of molecules and materials with tailored properties is challenging, as candidate molecules must satisfy multiple competing requirements that are often difficult to measure or compute. While molecular structures, produced through generative deep learning, will satisfy those patterns, they often only possess specific target properties by chance and not by design, which makes molecular discovery via this route inefficient. In this work, we predict molecules with (pareto)-optimal properties by combining a generative deep learning model that predicts three dimensional conformations of molecules with a supervised deep learning model that takes these as inputs and predicts their electronic structure. Optimization of (multiple) molecular properties is achieved by screening newly generated molecules for desirable electronic properties and reusing hit molecules to retrain the generative model with a bias. The approach is demonstrated to find optimal molecules for organic electronics applications. Our method is generally applicable and eliminates the need for quantum chemical calculations during predictions, making it suitable for high-throughput screening in materials and catalyst design.

## 1 Introduction

The search for new functional molecules and materials is often complicated by several criteria that must be simultaneously satisfied. For example, molecular materials tailored for organic electronics devices must be mechanically flexible, durable, and synthetically accessible while satisfying the relevant described electronic properties that govern the device functionality.[1,2] In addition to these often-competing requirements, it is not always clear how to systematically modify a molecular structure and composition to improve (multiple) properties. Simultaneous multi-property optimization can be considered the holy grail in molecular and material design.[2-4] A better understanding of how functional groups in a molecule alter its physicochemical properties could, at least in principle, help to facilitate design studies. However, the combinatorial complexity of chemical space consisting of up to $10^{60}$ organic molecules and the many factors that must be considered often make this problem too complex for traditional optimization and basic heuristic reasoning.[2,5] Candidate identification based on simple structure-property relations and trial-and-error optimization remain the state-of-the-art when it comes to developing new molecules and materials with specific property requirements.[2,6]

One area of research where this problem has become apparent is the field of organic optoelectronics, which deals with devices that emit or detect light. Examples in which novel organic electronic materials play a role range from sustainable energy sources (solar cells), organic light-emitting diodes (OLEDs), telecommunications, displays in smart devices, or optical fibers – to name a few examples. Organic thin film devices, composed of multiple organic layer components with different tailored properties have become of particular importance to this research area.[7,8] To deliver new molecular materials for thin film devices, their electronic properties such as the fundamental gap, the electron affinity, or ionization potential, must lie within a narrow window to satisfy the requirements of the device function.

Recently, generative deep learning has emerged as a promising solution for speeding up molecular design.[2,9-11] Generative deep learning is an unsupervised learning technique, in which deep learning models extract knowledge from a data set of (molecular) geometries and apply the acquired rules to create new molecules with properties similar to those in the original data set.[2] Several recent works have shown that such methods have the potential to dramatically accelerate molecular and material discovery[2,3,9,11-14], however, there is no guarantee that the generated molecular systems will exhibit properties within a relevant regime.

Unguided search in chemical space is extremely inefficient and fundamentally limits the diversity of structures that can be explored in high-throughput screening, particularly if the molecular generation process requires computationally demanding quantum chemical predictions of electronic properties. Even with hypothetically limitless computational resources, the characterization of generated molecules remains challenging. Several recent works have proposed property targeted generative workflows in the context of drug and molecular design. [15-18] Most generative models predict molecules via fragment-based structural descriptors such as SMILES strings that do not resolve the three-dimensional structure and conformation of molecules. Molecular generation can be guided by recursive workflows that use experimental reference data or quantum chemical calculations. In the latter case, the three-dimensional atomic configuration of the molecular equilibrium structure is required as input for quantum chemical calculations. Generative models that predict three-dimensional conformations of molecules have recently been proposed,[3,19-22] yet the requirement of performing quantum chemical calculations introduces a bottleneck that limits the number of molecules that can be screened.

In this work, we propose an approach that delivers high-throughput guided search and design of functional organic molecules with tailored properties. The method achieves this by combining two machine learning algorithms. The first model is an unsupervised, generative autoregressive model that can use chemical rules learned from a structural distribution of molecules to create new, previously unknown three-dimensional equilibrium conformations of molecules. The second model is a supervised physics-inspired deep neural network that, given a three-dimensional structure, can predict the (charged) electronic excitations of functional organic molecules with close to experimental accuracy.[23] The latter eliminates the need for demanding quantum chemical calculations used in previous approaches. The approach presented here provides an automated workflow in which chemical space exploration can be biased towards the generation of molecules that satisfy preset design parameters. We demonstrate the ability to perform high-throughput property-guided molecular design in the context of organic electronics.[24] Key molecular properties relevant in optoelectronic materials targeted here are small fundamental gaps, small ionization potential, and large electron affinity.[24,25] Important molecular features that separate the most optimal molecules with small fundamental gaps from the rest of the explored molecules can be unveiled using dimensionality reduction techniques and unsupervised clustering algorithms. The trends that we find and the rules we discover are verified with quantum chemistry, showing the potential of our method to discover hidden patterns in data. Finally, we provide an outlook for multi-property optimization by simultaneously biasing the generative model towards systems with low fundamental gaps and low synthetic complexity.[26]

**2      Results**

**2.1 Workflow**

The proposed approach for automated molecular design is a combination of two deep learning techniques, illustrated in **Figure 1**a. The process starts with training of a generative model on a set of molecular structures to learn underlying rules for building molecules that satisfy the same structural distribution and resemble the learned chemical space. The initially trained generative deep learning model is then used to predict a large number (in the range of several thousands to millions) of new molecules. A validity check of molecular structures is carried out and systems are filtered according to their structures, for example, duplicates or disconnected systems are discarded. For the structure generation, we use the generative, autoregressive deep neural network, G-SchNet.[3] G-SchNet, different to most other generative models,[2,27] is able to predict the structural composition and the three-dimensional conformation of molecules, which can serve as an input for electronic structure calculations and deep learning models of electronic structure.

The screening of molecular properties is facilitated with the deep neural network SchNet+H[23] to allow for high computational efficiency. SchNet+H predicts electronic excited states from equilibrium geometries that can be used to compute photoemission spectra with accuracy close to experiment. The high fidelity of the model is achieved by combining a deep learning model for molecular orbital energies obtained from density functional theory (DFT) and a Δ-ML model,[28] meaning that the difference (Δ) between two levels of theory is learned, to correct these energies to the accuracy of many body perturbation theory at the level of the GW method in the complete basis set limit. The GW method acts as a correction to DFT to account for many-body correlation and exchange effects. [25] As **Figure 1**a shows, molecules can also be screened based on other properties. In this work, molecules are additionally screened using another deep neural network capable of predicting the synthetic complexity of molecules, namely the SCScore,[26] which was trained on 12 million reactions from the Reaxys database.[29] The most promising molecules with properties that lie within a predefined target range are then used to bias the generative model, which can subsequently predict new molecules with electronic properties closer to the target.[3,12,30] By iteratively biasing the generative model, the properties of the predicted molecules can be pushed into unexplored regions.

We demonstrate the proposed workflow by training G-SchNet on the OE62 data set of functional organic molecules. The OE62 data is composed of molecules with large chemical and structural diversity. [24] As can be seen in **Figure 1**b, molecules can contain up to 16 different elements. They vary in size from 3 atoms to over 150 atoms. The distributions of fundamental gaps (ΔE) ionization potentials (IP), and electron affinities (EA) of molecules in the OE62 data set and of molecules generated by G-SchNet are shown in **Figure 1**c. ΔE, IP, and EA, are important measures to characterize molecules applicable in organic electronic devices and especially molecules with small ΔE are interesting and often used in photonics or biomedicine,[8] for instance. However, in the OE62 data set, there are not many molecules that exhibit small values of ΔE and IP or large values of EA in regimes that are typically considered relevant for organic electronics applications. Here, we demonstrate that, by iteratively biasing G-SchNet towards the desired property range, molecules can be designed that exhibit values of ΔE, IP and EA that lie outside of the property distribution represented by the original training data set.

**2.2 Biasing towards desired electronic properties**

The results obtained by iteratively biasing G-SchNet towards small ΔE, large EA, and small IP are shown in **Figure 2**. Panels a, c, and e show the distribution of targeted electronic properties for a set of 40,000 to 90,000 predicted molecules in each iteration (see **Supplementary Data 1**). In the first biasing step of each experiment, a small subset of molecules in the OE62 data with property values below or above a certain threshold (illustrated with shaded areas, which is about 10% of the OE62 data set) are used to retrain G-SchNet with a bias. The molecules are screened using SchNet+H, which has been independently validated to accurately predict electronic properties of structures predicted by G-SchNet in **Supplementary Section 2** of the supplementary information. SchNet+H has a mean absolute error for charged electronic excitations in the range of 0.25 eV with respect to the quantum chemistry reference, which we deem sufficiently accurate for high-throughput screening and identification of candidate molecules. To additionally ensure that G-SchNet is not misled by molecules that are inaccurately predicted with SchNet+H and thus wrongly assumed to fall into the category of molecule properties in the desired range, the variances of the electronic excitations inferred by two SchNet+H models are computed on-the-fly. Whenever the prediction variances are larger than their average mean absolute errors, we deem SchNet+H to be unreliable and the molecule is discarded, see Supplementary Section 3 for details.

As can be seen from **Figure 2** a, c, and e, after biasing and retraining of G-SchNet, molecules with a distribution of properties shifted towards the desired energy ranges can be generated. Interestingly, already after the first biasing steps, generated molecules exhibit electronic properties that lie outside of the original data set. In the subsequent iteration, the molecules with properties at the edges of the distributions are extracted and used to retrain G-SchNet again. The exact number of molecules and the criteria to select molecules used for biasing are specified in **Supplementary Section 4** and **Supplementary Data 1**. The molecular design process is terminated when the distribution of properties of proposed molecules as predicted by SchNet+H did not overlap anymore with the original distribution. This was after 7, 10, and 11 loops in the cases of ΔE, EA, and IP, respectively.

To verify that the distributions outside the original data set are not artefacts due to molecules outside of the training regime of SchNet+H, we recalculated ΔE, EA, and IP at the G0W0 level of theory for 66, 79, and 33 molecules, respectively, that are extracted randomly from the last 3 iterations of each experiment. Indeed, we found that the molecules consistently had electronic properties that were not present in the original data set. The smallest ΔE value of the extracted molecules is 3.2 eV, which is 1.6 eV smaller than the smallest value reported in the original data set. The largest EA is 6.6 eV, which is 2.4 eV larger than the largest EA reported in the original data set, and the smallest IP is 4.2 eV, 0.8 eV smaller than the smallest IP reported. The reference calculations and distribution of molecular properties are discussed in more detail in **Supplementary Section 2** (**Supplementary Figure 3**).

The fact that the generative model can produce molecules with ΔE, EA, and IP values that are not reported in the original training set may seem surprising, but this can be explained by taking a look at the chemical space spanned by the molecules in the OE62 data set and the structures predicted by G-SchNet (**Figure 1**b, d, and f). The chemical space represented and formed by the OE62 dataset is shown by two representative collective variables obtained from dimensionality reduction *via* principal component analysis (PCA) based on the smooth overlap of atomic positions (SOAP) structural descriptor[31] (see Methods section on Dimensionality reduction of generated molecules for details). The molecules generated in each consecutive biasing loop are shown in the same chemical space and indicated by different colors. As can be seen, the generated molecules are contained within the structural space spanned by the original OE62 data set. Interestingly, similar regions of chemical space are identified to be important for molecules that feature a small ΔE (panel b) and large EA (panel d). In contrast, a different region of chemical space is identified for molecules that predominantly exhibit a small IP (panel f).

**2.3 Identification of bonding patterns**

To correlate key bonding patterns in molecules with trends in electronic properties, we combined dimensionality reduction with clustering techniques and analyze which molecules feature small ΔE. For dimensionality reduction, PCA is applied to two types of descriptors, one that encodes bonding patterns and one that encodes structural distributions of molecules in the OE62 data set and for the collection of all molecules generated during consecutive biasing iterations. Five principal components obtained from each descriptor type were used as an input for clustering analysis, which contained over 98% of the variance in the data. Further principal components were not required as they would have negligible contribution to the analysis. For clustering, we used BIRCH[32] to find cluster centroids coupled with agglomerative clustering.[33] Details on descriptors, PCA, and clustering analysis can be found in the methods section. Data points plotted along the first principal components obtained from the structural descriptor against ΔE are shown in **Figure 3**, where colors in panel a indicate iterations and colors in panel b indicate subclusters found across iterations.

Manual inspection of the centroids of the subclusters indicated that an increased number of cyano groups (-C≡N) is present in molecules with small ΔE. This trend can also be observed for representative molecules plotted next to panel b of **Figure 3** and is quantified in **Figure 3**c. While in the original data set, mostly C-N single bonds are present and only few molecules have C≡N triple bonds, molecules generated during the last loops mainly contain C≡N triple bonds. To analyze whether this trend is sensitive to the original training data set, meaning to find out whether G-SchNet still predicts a high content of cyano groups in molecules optimized for small ΔE even if they are not contained in the original data, we eliminated all C≡N triple bonds from the original training data set and performed a knock-out study. The modified OE62 data set was used to train another G-SchNet model, which was then applied in a separate experiment to generate molecules with iteratively smaller ΔE. Already after the first loop, G-SchNet based on the knock-out data set generates molecules with an increased number of C≡N triple bonds (**Supplementary Figure 9**). The reason that G-SchNet can recover some functional groups not contained in the original data set lies in the nature of the SchNet descriptor, which is represented by a set of continuous atom-centered filter functions trained to optimally represent the data. These functions encode the probability of finding atoms at a certain distance and constellation around each atom. The G-SchNet model thus has some likelihood of also generating shorter CN bonds with different coordination than there are in the training set, which are then enhanced in the distribution via the biasing approach with SchNet+H.

Further, quantitative analysis of elemental composition of molecules generated by later loops revealed that a significant increase of sulfur and selenium content is found in molecules with small ΔE. The respective percentages are depicted in panels d-g of **Figure 3**. As can be seen from panel g, while sulfur and selenium content rises, the oxygen content decreases, which indicates replacement of oxygen by sulfur or selenium. To investigate whether this result is an artefact of our models or a real trend that leads to small ΔE, we carried out 144 quantum chemical calculations (see methods section on quantum chemistry calculations for details) of molecules with oxygen atoms replaced by sulfur and selenium and compared their HOMO-LUMO gaps as approximate analogues of fundamental gaps.[23,24] Our results clearly indicate that replacing one or all oxygen atoms with sulfur reduces the HOMO-LUMO gap on average by 0.5 eV and 1.1 eV, respectively. Further replacement of sulfur by selenium additionally decreases the HOMO-LUMO gap by 0.2 eV in both cases, hence, on average a decrease in HOMO-LUMO gap by 0.7 eV and 1.3 eV can be found when replacing one or all oxygen atoms with selenium atoms. The effect of selenium to promote photo-conducting properties was already reported in 1873.[34]

Molecules predicted by G-SchNet contain unusually high concentrations of selenium and sulfur atoms as well as cyano groups considering that they were generated from a base distribution of known crystal forming organic molecules. To find out if such molecules are used in real applications, a literature search with SciFinder[35] was conducted (**Supplementary Figure 8**). In addition to literature search, we parsed all molecules from the final 3 loops and compared them with approximately 250k small aromatic molecules considered applicable to organic electronics[36] by using the Tanimoto similarity measure[37] (see Methods section on Similarity analysis of molecules for details). The findings suggest that the identified molecules contain structural motifs, such as tetrathiafulvalenes[38] and (selenium-enriched) tri-thiapentacene derivatives[39] shown in bold in panel b, that are frequently mentioned in literature relevant to organic molecular electronics,[40] especially in the context of (dye-sensitized) solar cells,[41,42] for synthesis of organic electronic materials, electroluminescent materials,[43,44] or single-molecule switches.

As it is evident from the results above and **Figure 3**c-g, G-SchNet changes the relative distribution of elements and bonding patterns to shift the electronic properties into the desired range of small ΔE. In doing so, molecules are generated that feature known structural motifs that are already in use in organic electronics.

However, the property-based biasing approach also comes with downsides. While the biased generative method successfully creates molecules with desirable properties, as iterations progress, the method also creates molecules with narrow structural distributions and increasing synthetic complexity and, in many cases, with highly improbable structural arrangements. The complexity of synthesizability of the generated molecules is shown in **Figure 3**h and obtained from a neural network for the SCScore metrics by Coley et al.[26]. The SCScore ranges from 1 (low synthetic complexity) to 5 (high synthetic complexity) and as can be seen, the minimization of ΔE comes at the detriment of the synthetic complexity of the molecules. Ideally, molecules should be designed that feature electronic properties in an optimal range while still being synthetically accessible.

**2.4 Targeting multiple properties**

To generate molecules that exhibit both low synthetic complexity and small ΔE, we selected 2670 molecules with small ΔE and small SCScore out of an initial data set created from merging the OE62 data set with an additional set of 340k molecules generated with G-SchNet trained on OE62. These data points were used to bias G-SchNet and in each consecutive loop, molecules were selected that satisfy selection criteria for both properties. The distributions of ΔE and SCScore for each iteration are shown in **Figure 4** a and b, respectively. As can be seen, after each biasing step, G-SchNet successfully predicts molecules with iteratively smaller ΔE and smaller SCScores. Analysis of the elemental distributions of the generated molecules (**Supplementary Figure 11**) reveals that the overall structural trends observed in single-property biasing of molecules that lead to small ΔE are retained. However, selenium is effectively eliminated from the distribution due to the additional criterion of achieving small SCScore. This trend is encouraging as selenium is a trace element and less abundant than sulfur. In addition, it is considered a contaminant of concern in water systems.[45] This is especially problematic as selenium has one of the narrowest windows between concentrations where it serves as a vital trace mineral and concentrations where it is toxic, hence industrially caused accumulation in the environment poses a risk.[45,46]

## 3     Discussion

The presented method constitutes an efficient workflow for the (multi)-property-driven design of previously unseen molecules. One of the limitations of the model is that it requires the prediction and screening of several hundred thousands of molecules in each loop to obtain a large enough number of molecules with which the generative model can be biased after screening. This process is limiting, especially when the chemical diversity of generated structures is small and can become a computational bottleneck if molecules are screened towards more than two properties. This limitation can be tackled with conditional generative models, such as conditional G-SchNet,[12] which enable the conditioning of the generative model towards predicting molecules with certain properties by including these properties of interest as labels during training.

The ability to generate viable molecules is not unique to G-SchNet and other previously proposed generative models have shown to achieve similar results. The novelty of our approach lies in its high-throughput capability. For example, previously reported approaches, such as the one by Sumita et al.,[16] perform generative search based on SMILES strings, which are translated into three-dimensional structures with RDKit[47] and then screened using quantum chemistry calculations. This has several downsides. Firstly, the conversion with RDKit of the generated structures does not necessarily yield equilibrium structures, whereas G-SchNet is only trained on relaxed equilibrium structures and was previously shown to predict structures close to structural equilibrium (see Supplementary Figure 1).[3] A prediction based on SMILES would also not have allowed us to predict cyano groups from an original training database that does not contain such functional groups. Furthermore, the screening of 1,000 generated molecules with quantum chemistry calculations at the accuracy that we require would have taken over 500,000 computing hours or roughly 20,000 days. In contrast, in this work, we have screened many hundreds of thousands of molecules in few days. The combination of the ML models applied here is thus a clear advantage that provides true high-throughput molecular design capabilities.

The ability of the method to predict molecules with electronic properties beyond the initial training data set will be useful for a range of applications from high-throughput drug discovery to molecular design for organic electronics. Future work will explore how the performance of the method can be further improved by using different neural network architectures. By coupling this approach with a generative model of condensed phase structures, the property-driven design of crystalline solids may be possible.

## Methods

**Quantum chemistry calculations**

Quantum chemistry calculations to verify results were carried out using the same procedure as in Ref.[24] that was used to generate the data set. All calculations were carried out using FHI-aims.[48] Every molecule was first relaxed using DFT with the PBE functional[49] and the standard default "light" basis set as defined in FHI-aims. We augment the PBE functional with the Tkatchenko-Scheffler (PBE+vdW) correction to account for long-range dispersion corrections.[50] Afterwards, structure relaxations using the same settings, but with a standard default "tight" basis set were carried out. PBE0[51,52] orbital energies were calculated based on the PBE+vdW optimized structures.

Using the PBE+vdW optimized structures, additional G0W0@PBE0 calculations were carried as implemented in FHI-aims with analytic continuation.[53] To extract quasiparticle energies in the complete basis set limit, two calculations were conducted: one with the triple-zeta basis set def2-TZVP and one with the quadruple-zeta basis set def2-QZVP.[54] The extrapolated values were calculated by a linear regression against the inverse of the total number of basis functions.[24,55]

To analyse the effect of sulfur and selenium content in molecules, we carried out DFT calculations of 144 randomly selected molecules generated with G-SchNet that contained no sulfur and no selenium, but oxygen atoms. We then carried out 5 calculations, one with the original molecule, two with a molecule in which a single oxygen atom is once replaced with a selenium atom and once with a sulfur atom and two with a molecule in which all oxygen atoms are replaced with either sulfur or selenium. The HOMO-LUMO gaps were compared as approximates to fundamental gaps, because despite them being underestimated with DFT, the trends are similar to those found with G0W0.[23,24]

**G-SchNet for OE62**

G-SchNet was originally developed for small organic molecules made up of carbon, hydrogen, oxygen, nitrogen, and fluorine (QM9 data set[56,57]). We adapted G-SchNet to train on molecules that are part of the OE62 data set,[24] which features large chemical and structural diversity (**Figure 1**b) and contains 62k molecular structures that are extracted from experimentally discovered organic crystals.

To generate molecules, the autoregressive, generative model learns from atomic positions, $r_i$, and corresponding atom types, $Z_i$, $\boldsymbol{R}_{\leq n} = (r_1, \ldots, r_n)$ with $r_i \in \mathbb{R}^3$ and $Z_{\leq n} = (Z_1, \ldots, Z_n)$ with $Z_i \in \mathbb{N}$, respectively. Thus, $n$ point sets of atom types and positions are considered.

Rotationally and translationally invariant feature vectors are created using SchNet,[58,59] a continuous-filter convolutional neural network that was originally developed to map molecular structures to properties like energies or polarizabilities. The atomic features obtained from SchNet are multiplied elementwise with outputs of an embedding layer obtained from atom types and two additional auxiliary tokens. The number of tokens can be generalized using the variable *t*. The resulting feature vectors are then processed using dense atom-wise layers to obtain the probabilities of the next atom types and positions. The probability of the next atom type, $p(Z_{t+i}|\mathbf{R}^t_{\leq i-1}, \mathbf{Z}^t_{\leq i-1})$, is obtained via:

$$p(Z_{t+i}|\mathbf{R}^t_{\leq i-1}, \mathbf{Z}^t_{\leq i-1}) = \frac{1}{\beta} \prod_{j=i}^{t+i-1} p(Z_{t+i}|x_j). \tag{1}$$

Probabilities for atomic positions of the next atom are obtained in a similar way. Note that due to *t* auxiliary tokens that can be seen like auxiliary atom types that do not belong to the final generated molecule, indices run from 1 to *t+n*. One token marks the origin of the structure generation process and is fixed. The use of this token was found to improve training and lead to generated structures closer to the original distribution. In addition, another token, meaning, the focus point, breaks the symmetry of molecules and reduces artefacts. Each additional atom is always placed such that it is a neighbor of the focus point. $\beta$ is a normalization constant. [3]

Note that to generate rotationally equivariant probabilities, the 3-dimensional information is obtained from pairwise distances, $d_{(t+i)j} = \|r_{t+i} - r_j\|_2$, rather than absolute positions, with $\alpha$ being a normalization constant.

$$p(r_{t+i}|\mathbf{R}^t_{\leq i-1}, \mathbf{Z}^t_{\leq i}) = \frac{1}{\alpha} \prod_{j=i}^{t+i-1} p(d_{(t+i)j}|\mathbf{R}^t_{\leq i-1}, \mathbf{Z}^t_{\leq i}) \tag{2}$$

To train G-SchNet on molecules of the OE62 data set, the original code was adapted. Importantly, the additional atom types that are present in the OE62 data set compared to the QM9 data set had to be added and minimum and maximum bonding distances and orders had to be defined. In addition, since only molecules with even numbers of electrons were available in the OE62 data set, we added a filtering function that excluded all molecules with unpaired electrons.

G-SchNet has the advantage of generating molecules in 3d. We validated that the generated molecules are close to their equilibrium structures according to the reference method, DFT in **Supplementary Section 1** of the supplementary information (SI).

G-SchNet for OE62 was trained using a batch size of 2, a cut-off of 10 Å, 128 features (size of atom-wise representation), 9 regular SchNet interaction blocks, and 25 Gaussian functions to expand distances between atoms. In G-SchNet, the batch size depends on the number of samples per batch, but also on the size of the molecules within a batch. This is, because a molecule is generated one atom at a time. Per default, the whole trajectory to create a molecule is sampled, which can lead to large memory consumption, especially when molecules in a batch are large. Since molecules in the OE62 data set can contain up to 200 atoms, we drew 5 random atom placements per molecule per batch instead of the complete trajectory. To still sample from the whole trajectory during training, a high number of epochs was chosen. Besides these, default parameters were used to train G-SchNet, which is an initial learning rate of 0.0001 and a decay of the learning rate by 0.5 after 10 epochs without improvement of the model during training.

**SchNet+H for quasiparticle energies**

The unsupervised, autoregressive generative deep neural network, G-SchNet,[3] is combined with a supervised, physically-inspired deep neural network to design molecules with decreasing ionization potential, IP, increasing electron affinity, EA, as well as decreasing fundamental gap, ΔE. Compared to recent studies that aimed to optimize the HOMO-LUMO (highest occupied molecular orbital-lowest unoccupied molecular orbital) gap as a theoretical proxy of the fundamental gap, we optimize the gap as obtained from charged electronic excitations.[60] We used the already-trained SchNet+H models from Ref. [23] for this study.

As illustrated in the bottom of **Figure 2**, the ionization potential of a given state *i*, $IP_i$, describes the energy of a bound state, which can be reconstructed experimentally in photoelectron spectroscopy by ejection of electrons with kinetic energy, $E_{\text{kin}}$, from a sample with work function, $\Phi$, after irradiation with UV/visible light or X-rays with energy, $h\nu$:

$$IP_i = h\nu - E_{\text{kin}} - \Phi = -\varepsilon_i \quad for \quad \varepsilon_i < E_{\text{Fermi}} \tag{3}$$

$E_{Fermi}$ indicates the Fermi level and $\varepsilon_i$ the electron removal energy or quasiparticle energy of ionization. In contrast, the electron affinity of a state *i*, $EA_i$, is equal to the negative energy of unoccupied states or the quasiparticle energy of electron addition and can be measured by measuring emitted Bremsstrahlung of electrons scattered in a sample:

$$-EA_i = E_{\text{kin}} - h\nu + \Phi = -\varepsilon_i \quad for \quad \varepsilon_i \geq E_{\text{Fermi}} \tag{4}$$

The fundamental gap is the energy difference between ionization potential and electron affinity. The HOMO and LUMO energy levels according to DFT are often used to approximate the IP and EA, respectively, because they are computationally cheaper to calculate, but less accurate compared to many body perturbation theory at the GW level of theory. Consequently, the HOMO-LUMO gap is often used as an approximation of the fundamental gap but is known to underestimate energies.[24]

In this work, the GW quasiparticle energies are obtained from SchNet+H, a physically inspired deep neural network trained on orbital energies from DFT/PBE0 of molecules in the OE62 data set. As in G-SchNet, SchNet+H uses the SchNet-descriptor[58,59] to represent molecules. In contrast to G-SchNet or the conventional SchNet model for molecular properties, however, SchNet+H predicts multiple energy levels, $\varepsilon_i^{\mathrm{ML(DFT)}}$, by inferring a latent Hamiltonian, $\boldsymbol{H}^{\mathrm{ML(DFT)}}$, which is diagonalized using a transformation matrix, $\boldsymbol{U}$:

$$diag\left(\left\{\varepsilon_i^{\mathrm{ML(DFT)}}\right\}\right) = \boldsymbol{U}^T \boldsymbol{H}^{\mathrm{ML(DFT)}} \boldsymbol{U}. \qquad (5)$$

In this way, a transferable representation of molecular energies for molecules of arbitrary sizes is created. The energy levels obtained after diagonalization of the ML-inferred Hamiltonian can be corrected to GW accuracy at the complete basis set limit by another model trained on the difference between $\varepsilon_i^{\mathrm{ML(DFT)}}$ and $\varepsilon_i^{\mathrm{GW}}$, meaning quasiparticle energies at the GW level of theory. Adding corrections to the energy levels, $\varepsilon_i^{\mathrm{ML(DFT)}}$, results in energy levels at the GW level of theory:

$$\varepsilon_i^{\mathrm{ML(GW)}} = \varepsilon_i^{\mathrm{ML(DFT)}} + \varepsilon_i^{\mathrm{ML(GW-DFT)}}. \qquad (6)$$

This model has been shown to be accurate to predict photoemission spectra of molecules in the OE62 data set and functional organic molecules outside of this data set.[23] In this work, this model is applied to screen G-SchNet-predicted structures based on their fundamental gap, electron affinity, and ionization potential. The applicability of SchNet+H for this purpose was validated in **Supplementary Section 2**.

**Computational details of the workflow for targeted design**

The generation of molecules with desired electronic properties was conducted by biasing G-SchNet, meaning retraining it, with a subset of molecules that exhibit specific electronic properties. In this work, G-SchNet was biased independently 3 times: towards small ΔE, large EA, and small IP.

In each loop, we generated 200k molecules for biasing towards small ΔE, large EA, and multiple properties and 100k for biasing towards small IP and during the knockout study. The number for IP biasing was reduced to keep the balance between computational effort and accuracy, as molecules generated during this loop were on average larger and required about twice the computational resources. As 100k molecules yielded satisfactory results, while reducing computation times, this number was selected for the knockout study too. One loop in all studiestook approximately 2 days. This time includes the molecule generation and the screening of these molecules with SchNet+H (computational costs of SchNet+H are specified in **Supplementary Section 5** in the SI). These molecules were then sorted based on their electronic properties. Those molecules with electronic properties (IP and ΔE) below their mean minus standard deviation or (EA) above their mean plus standard deviation were selected for re-training of G-SchNet.

When biasing towards multiple properties, which are the SCScore[26] and ΔE, we found that, out of the OE62 data set, only 47 of the predicted molecules had lower fundamental gaps and lower SCScore than their respective means minus standard deviation. To increase the initial data set for biasing, another 340k molecules were generated with G-SchNet and molecules with values for SCScore and ΔE lower than their mean minus 0.5 times standard deviation were selected, which resulted in an initial biasing data set of about 2670 molecules. During every biasing step, molecules that had fundamental gaps smaller than their mean minus 0.5 times standard deviation and SCScore small than their mean minus 0.5 times standard deviation or SCScore ≤ 2, were selected for biasing G-SchNet in the next iteration.

For biasing towards small IP, the process terminated after 2 loops due to the generation of very large molecules, which made the structure generation and filtering process extremely computationally costly and finally, infeasible with the existing computational resources at the time. As stated in Ref. [12], where a conditional G-SchNet model was trained on drug-like molecules with about 50 atoms at most, further adaptions, such as a cutoff or long-range interactions, are needed to allow for scalability to larger systems. The problem was circumvented in this work by restricting the IP-biasing experiment to the prediction of molecules with up to 70 atoms. With this adaption, the biased retraining could be conducted straightforwardly.

We terminated each experiment by continuously checking the electronic properties of a predicted data set and their chemical diversity. As soon as the distribution of properties for the predicted molecules did not significantly change between iterations the workflow was terminated. All loops until then were used for analysis. We ended up with 7, 11, and 10 iterations for biasing towards small ΔE, small IP, and large EA, respectively.

**SCScore to predict the complexity of synthesizability of molecules**

To estimate the complexity of the synthesis of a molecule, the SCScore is used as obtained from a deep neural network trained on 12 million reactions from the Reaxys data base.[29] This score correlates with the number of steps used for synthesis. As inputs, this model uses canonical SMILES strings[61] that are generated using Open Babel.[62]

The SCScore runs from 1 to 5, whereas molecules that have an SCScore of 5 are expected to be highly complex to synthesize and molecules with an SCScore of 1 are expected to be easily synthesizable. The SCScore defines synthesizability according to the number of reactions steps that are needed to synthesize from reasonable starting materials. Information on what starting materials are useful is learned and thus included in the model implicitly.[26]

**Dimensionality reduction of generated molecules**

To visualize the chemical space that is spanned by molecules generated with G-SchNet compared to molecules in the original OE62 data set and to create inputs for subsequent cluster analysis, we applied dimensionality reduction, for which we used principal component analysis (PCA) as implemented in scikit-learn.[63]

The inputs for PCA were one of two applied molecular descriptors that we refer to as bonding and structural descriptors. The structural descriptors are obtained using the smooth overlap of atomic positions (SOAP) descriptor[31] that leads to a 57,792-dimensional description of molecules with the aim of accounting for the whole molecule. To obtain bonding descriptors, we take raw molecular geometries and apply OpenBabel[62] and RDKit[47] to extract as many interesting features relating to the bonding of the molecules as possible. Features can be as simple as the number of atom types in a molecule, but also more complex, such as the number of rings of certain sizes, the aromaticity of a molecule, and targeted electronic properties. The final dimension of the bonding descriptor was 732.

Each defined descriptor obtained from molecules of the OE62 data set was used as input for PCA. To visualize the chemical space spanned by the OE62 data set in comparison to the space spanned by the G-SchNet-generated molecules, we generated the descriptor for G-SchNet-generated structures and represented them using the same principal components as obtained from the OE62 data. The first two principal components cover 94% and 90% of the variance in the OE62 data set for the bonding and structural descriptor, respectively (see **Supplementary Figure 5**). Results obtained from bonding descriptors are shown in the SI in **Supplementary Figure 6**.

**Clustering analysis**

For clustering, we use a mixture of Birch[32] and agglomerative clustering[33] to allow for uneven cluster sizes as implemented in scikit-learn,[63] which was chosen due to its memory efficiency. As an input, we used 5 principal components obtained after PCA of all molecules using both previously defined descriptors. The inputs were normalized such that features obtained from the different descriptors are equally weighted. An exception to this is that clustering was weighted towards changes in energy to better resolve energetic trends in subclusters. Clustering was conducted for molecules pooled from all biasing iterations. For each of the 13 clusters found, we extracted 10 molecules around the centroid. This procedure provides us with a condensed view into what makes up each cluster and reduces the task of analyzing clusters that contain too many molecules for manual inspection. The clusters found are illustrated using structural principal components and small fundamental gaps in **Supplementary Figure 7** a-b and the subclusters are shown in panel c of the same figure.

**Similarity analysis of molecules**

To measure similarity of molecules generated with G-SchNet and found in literature, we used SMILES strings and computed the Tanimoto score.[37] The mean of the Tanimoto score between molecules obtained after biasing against small fundamental gaps is 0.54. Note that a similarity of 0.5 is often considered significant.[64] The maximum similarity found was 0.72. We used key groups of molecules that exhibited the highest similarities and searched for hits using SciFinder (**Supplementary Figure 8**).

# Data availability

The OE62 data set is available in ref. [24] and the OE62+340k G-Schnet molecule dataset is uploaded on https://figshare.com/articles/dataset/G-SchNet_for_OE62/20146943.[65] Quantum chemistry calculations carried out in this study are uploaded to NOMAD under DOI: 10.17172/NOMAD/2022.07.02-1.[66] A supplementary data file showing the number of molecules that were predicted and used for training in each experiment and each loop is included as Supplementary Data 1. Source data for Figures 1-4 is available with the manuscript.

# Code availability

The modified G-SchNet version is available on github: https://github.com/rhyan10/G-SchNetOE62 and tagged as version v0.1 (minted version under DOI: 10.5281/zenodo.7430248).[67] The github repository includes scripts to analyze the data and carry out PCA. SchNet+H is published in Ref.[23] and available on http://www.github.com/schnarc (minted version under DOI: 10.5281/zenodo.7424017).[68] We include a tutorial for using SchNet+H and G-SchNet models for OE62 on figshare: https://figshare.com/articles/dataset/G-SchNet_for_OE62/20146943 including instructions for installation.[65] Original tutorials for training and using G-SchNet and SchNet+H are available on the github of the original code of G-SchNet https://github.com/atomistic-machine-learning/G-SchNet)[3] and SchNarc (https://github.com/schnarc/SchNarc/tree/develop),[69] respectively.

## Acknowledgements

This work was funded by the Austrian Science Fund (FWF) [J 4522-N] (JW), the EPSRC Centre for Doctoral Training in Modelling of Heterogeneous Systems [EP/S022848/1] (RJM), the EPSRC-funded Network+ on Artificial and Augmented Intelligence for Automated Scientific Discovery [EP/S000356/10] (RJM), and the UKRI Future Leaders Fellowship program [MR/S016023/1] (RJM). Computational resources have been provided by the Scientific Computing Research Technology Platform of the University of Warwick, the EPSRC-funded Northern Ireland High Performance Computing service [EP/T022175/1] via access to Kelvin2, the EPSRC-funded HPC Midlands+ computing service [EP/P020232/1] via access to Athena and Sulis, and the EPSRC-funded High End Computing Materials Chemistry Consortium [EP/R029431/1] for access to the ARCHER2 UK National Supercomputing Service (https://www.archer2.ac.uk). The authors thank Niklas Gebauer (TU Berlin) for fruitful discussions on the GSchNet model. This version of the article has been accepted for publication, after peer review (when applicable) but is not the Version of Record and does not reflect post-acceptance improvements, or any corrections. The Version of Record is available online at: http://dx.doi.org/10.1038/s43588-022-00391-1.

## Author Contributions Statement

R. J. M. conceived the original idea and supervised the research project. R. J. M. and J. W. designed the research project. R. B. and J. W. trained the deep learning models and created the property-guided design workflow. J. G. and J. W. performed the data set curation, predictions, model validation and data analysis. J. W. performed the quantum chemistry calculations. J. W. and R. J. M. wrote the manuscript with the help of the other authors. The manuscript reflects the contributions of all authors.

## Competing Interests Statement

Reinhard Maurer is an editorial board member of the journal Communications Materials. All other authors declare no competing interests.

## Figure Legends/Captions:

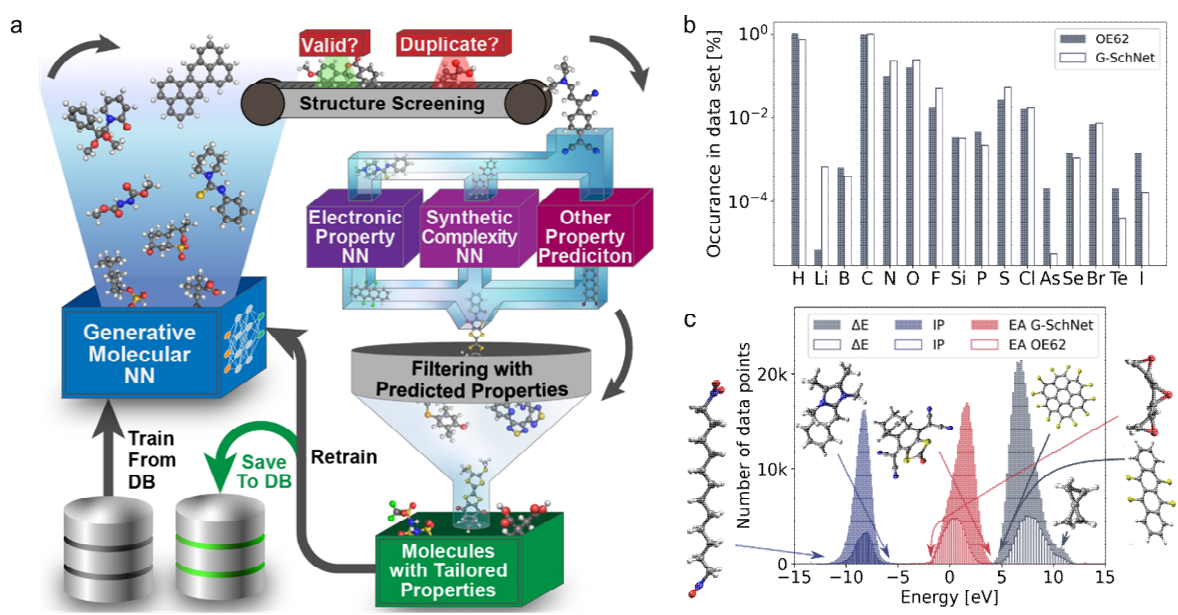

*Figure 1: Workflow of the proposed method and distribution of molecules in the data set. a) The proposed method starts by training the generative deep learning model, G-SchNet, on the OE62 data set, which can then be applied to build three-dimensional conformations of unseen molecules. These are filtered based on structure, for example, duplicates or disconnected structures are sorted out, and based on electronic properties, synthesizability, or other properties. In this work, we use SchNet+H to screen for small fundamental gaps, small negative electron affinity, and large ionization potential. In addition, we apply the SCScore neural network (NN) model to screen for molecules with low complexity in synthesizability. Selected molecules can be used to retrain and bias the generative model. b) Elemental distribution of molecules in the OE62 data set and those predicted by G-SchNet. c) The distribution of the fundamental gap, ΔE, ionization potential, IP, and electron affinity, EA, in the OE62 data set and in molecules predicted with G-SchNet. Example molecules are shown in the plot that highlight chemical and structural diversity of molecules in the data set. DB refers to data base in the image.*

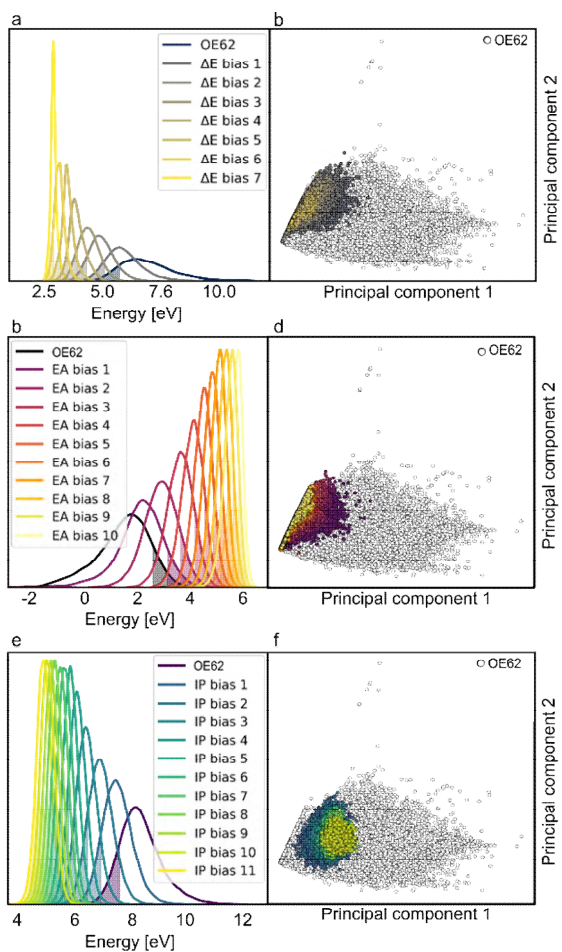

*Figure 2: Distribution of electronic properties and structures of generated molecules.* Distribution of fundamental gap, $\Delta E$, (a), electron affinity, EA, (c), and ionization potential, IP, (e) after biasing towards small $\Delta E$ (a), large EA (c), and small IP (e) b,d,f) Distribution of data points in chemical space spanned by principal components obtained from OE62 data (white circles) using structural descriptors. The color code indicates the biasing step.

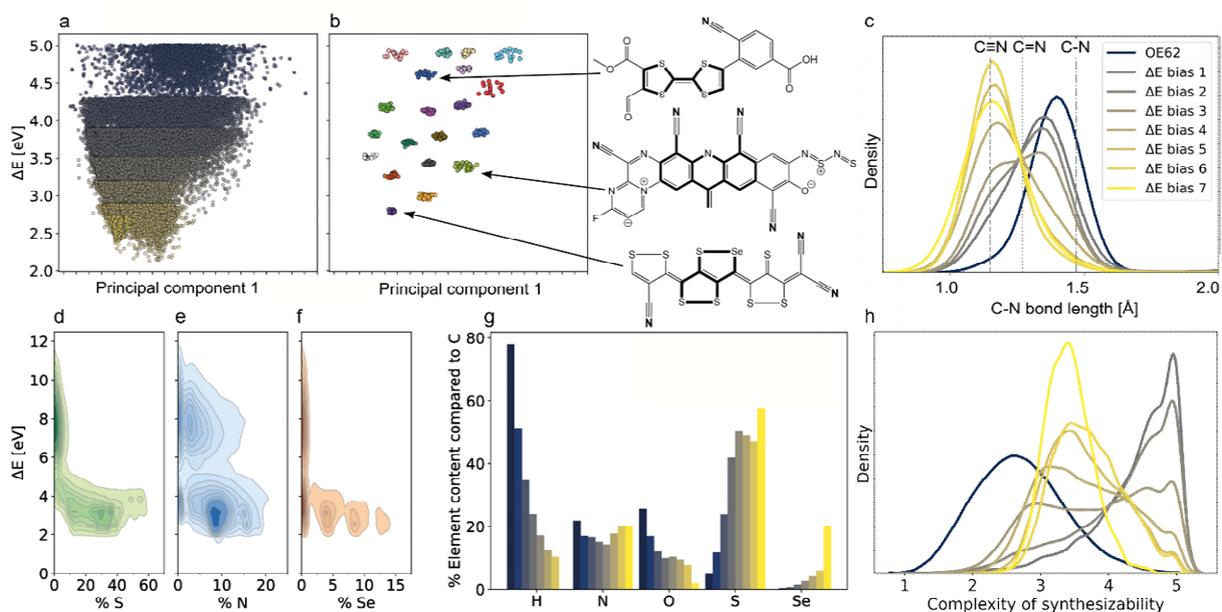

*Figure 3: Cluster analysis for molecules with small fundamental gaps. a) Molecules obtained with G-SchNet after biasing towards small fundamental gaps are represented using the first principal component using structural (SOAP) descriptors of all molecules (OE62 and G-SchNet-generated molecules). The color gradient corresponds to the different loops. The same legend as in panel c applies. b) Subclusters found with unsupervised learning obtained from data of a) and representative molecules illustrated next to it. c) C-N bond length distribution. Relative elemental content of d) sulfur (S), e) nitrogen (N), and f) selenium (Se) in molecules obtained with G-SchNet and in the original data set. g) Elemental composition and h) distribution of the synthetic complexity score (SCScore) of molecules of the OE62 data set and obtained from G-SchNet (the same legend as in c) applies).*

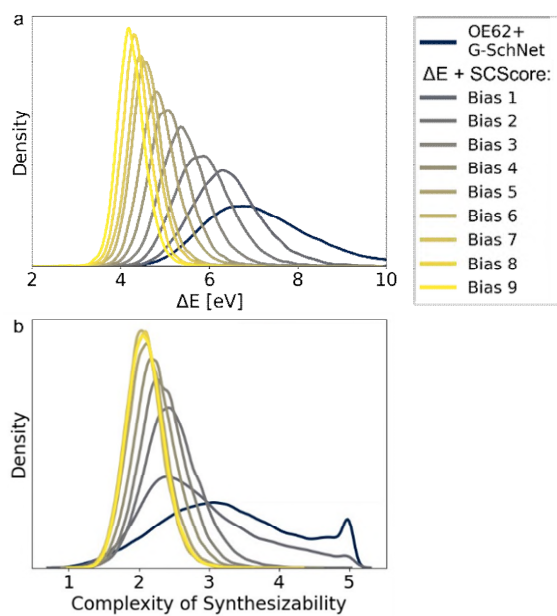

**Figure 4: Multi-property biasing.** *a*) Distribution of fundamental gaps, ΔE, and b) synthetic complexity score (SCScore) of molecules in the OE62 data set and generated with G-SchNet biased against both properties.

# Table of Contents

**Supplementary Section 1      Validation of G-SchNet for OE62**

To ensure that the trained autoregressive, generative deep neural network, G-SchNet,[1] predicts sensible structures that resemble the molecules in the original data set (OE62),[2] we carried out a two-fold analysis. First, we generated a data set of 100k molecules with G-SchNet with molecules that contain up to 100 atoms. We then randomly selected 400 data points and optimized them with PBE+vdW[3-5] and tight basis set settings using FHI-aims.[6] For structure relaxations, the same protocol reported for the OE62 data set was used (see also Methods section on quantum chemistry calculations).

The optimized molecular geometries were then aligned with the G-SchNet predicted molecules and the root mean squared deviations (RMSD) were computed. The distribution of RMSD values is shown in **Supplementary Figure 1**. In addition to the RMSDs, sample molecules are shown. The G-SchNet predicted structures are solid, while the density functional theory (DFT)-optimized structures are shown slightly transparent. The left-most molecule shows the structure with lowest RMSD of 0.022 Å, where both structures are almost identical. The second pair of two structures illustrates deviations of around 0.5 Å (i.e., 0.46 Å and 0.48 Å) and deviations are representatives for most of the predicted molecules with G-SchNet. The next pair of two structures to the right have an RMSD at around 1.09 Å und 1.18Å. At RMSD > 1Å, deviations between G-SchNet-predicted structures and DFT-optimized structures become clearly visible but can be deemed minor. The molecule with the largest deviation of 3.00 Å is shown on the right and shows a geometry that GSchNet predicts to be more strongly distorted than the DFT reference result.

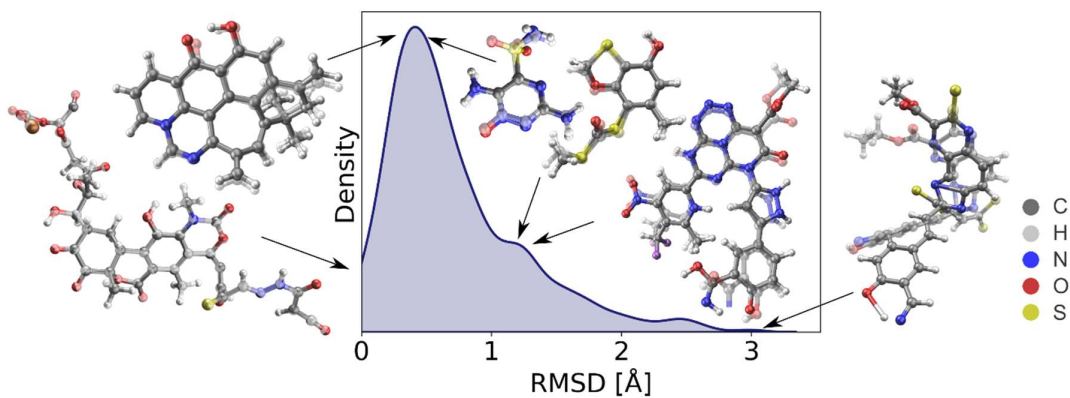

*Supplementary Figure 1: Validation of G-SchNet predicted structures. The root mean squared deviations (RMSD) of molecules predicted with G-SchNet were compared to structures obtained after structure relaxation with the reference density functional theory method. Exemplary molecules are shown, where the G-SchNet predicted structure (solid colors) is overlaid with the DFT-optimized structure (transparent). Examples for molecules that have very low RMSD, RMSD at around 0.5 Å, 1.1-1.2 Å, and >3 Å are illustrated.*

In addition to the RMSD, we compared distributions of some the most common bond lengths and bond angles. This analysis is based on the validation of G-SchNet that was carried out for the QM9 data set in Ref.[1]. The distributions for C-C-C bond angles, C-O-C bond angles, C-O bond distances, C-C-O bond angles, C-H bond distances, and ring sizes of molecules in the OE62 data set and molecules predicted by G-SchNet can be seen in **Supplementary Figure 2**. As can be seen, the distributions are very similar and indicate that, at least for the illustrated bonds and angles, G-SchNet structures resemble the molecular structures of the OE62 data set. The similarity of G-SchNet structures compared with molecules of the OE62 data set can be further assessed from Figure 1b, which shows the elemental composition of molecules with respect to the amount of carbon. Besides the amount of lithium and arsenic, which appear to differ strongly in the plot but in reality deviate only minorly due to the log-scale used for better visibility of elements with negligible amounts, the molecular compositions are very similar.

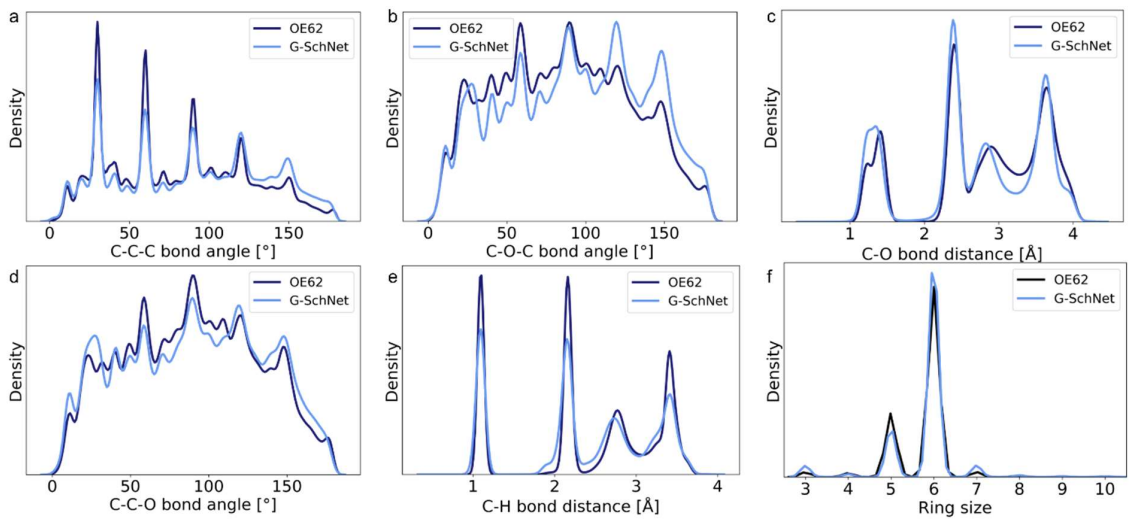

*Supplementary Figure 2: Comparison of structures predicted with G-SchNet with structures of the OE62 data set. a) Probability distribution of C-C-C bond angles, b) C-O-C bond angles, c) C-O bond distances, d) C-C-O bond angles, e) C-H bond distances, and f) ring sizes of molecules in the OE62 data set and G-SchNet-predicted molecules.*

## Supplementary Section 2     Validation of SchNet+H for G-SchNet-predicted structures

To assess the influence of structural differences on the electronic properties of molecules, i.e., orbital energies and quasiparticle energies, we predicted orbital energies of the 400 molecules used for validation of G-SchNet in **Supplementary Figure 3**, as obtained from G-SchNet and after structure optimization with DFT. The orbital energies of DFT-optimized structures predicted with SchNet+H are plotted against orbital energies obtained from DFT (PBE0[7,8] and tight basis set settings) using the DFT-optimized molecules as inputs. The mean absolute error (MAE) is about 0.24 eV (**Supplementary Figure 3**a). For comparison, the error of SchNet+H orbital energies for G-SchNet predicted structures compared to orbital energies obtained with DFT using DFT-optimized structures are only slightly larger, i.e., 0.26 eV (**Supplementary Figure 3**b). The same test was executed with SchNet+H for quasiparticle energies. The MAE error obtained using DFT-optimized and G-SchNet predicted structures is about 0.25 eV and 0.28 eV, respectively. Scatter plots of quasiparticle energies using G-SchNet-predicted structures for SchNet+H predictions can be seen in **Supplementary Figure 3**c. For comparison, the error of SchNet+H for molecules in the test data of the OE62 data set is about 0.13 eV. The method can be deemed sufficiently accurate for the purpose of high-throughput targeted design of functional organic molecules.

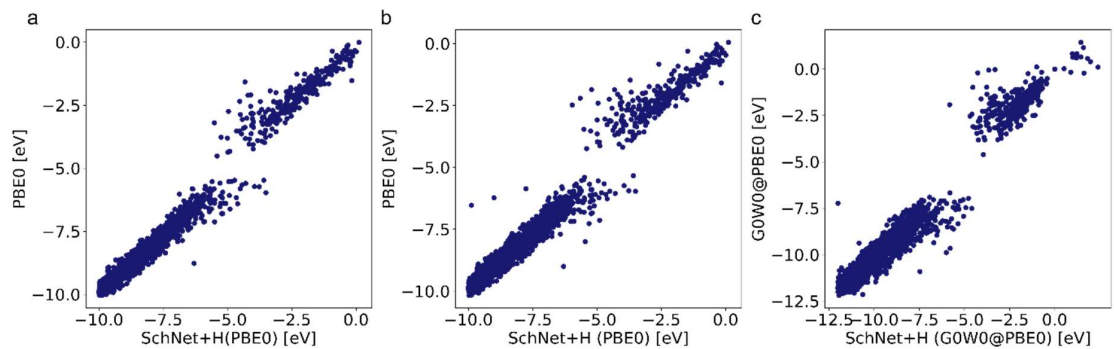

*Supplementary Figure 3: Validation of SchNet+H for G-SchNet generated structures. **a)** Scatter plots of SchNet+H predicted orbital energies and PBE0 orbital energies for structures obtained from G-SchNet and for **b)** optimized structures with PBE+vdW and the tight basis set settings. The same procedure as in the original data set was carried out to relax molecules. **c)** Scatter plots of SchNet+H predicted quasiparticle energies and reference G0W0@PBE0 quasiparticle energies for molecules obtained from G-SchNet without additional DFT optimization.*

**Supplementary Section 3    Validation of molecules at the edges of of the distributions**

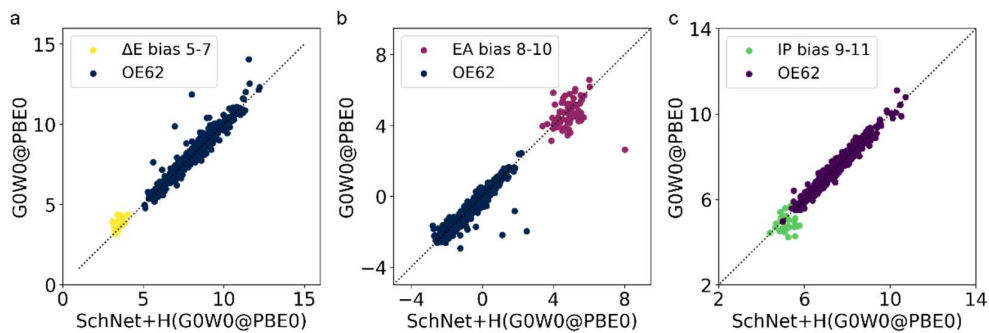

*Supplementary Figure 4: Validation of electronic properties of generated molecules. **a)** Fundamental gaps, ΔE, **b)** electron affinities, EA, and **c)** ionization potentials, IP, for molecules of the original data set and G-SchNet generated structures of the last 3 biasing steps predicted with SchNet+H and computed with G0W0@PBE0.*

To assess the reliability of SchNet+H predictions, two SchNet+H models were employed that were trained on different random train/test splits of the same dataset. By computing the deviation of ΔE, EA, and IP values between the two models, which should be well below the MAE of the individual models, it is possible to identify structures for which quasiparticle energies are predicted with high uncertainty. The threshold was set to the MAE of the models that was determined for a given data set. This approach is known as query by committee. [9,10]

To further validate the predictions of SchNet+H for molecules obtained in the last biasing steps of the fundamental gap, ΔE, the electron affinity, EA, and the ionization potential, IP, G0W0@PBE0 calculations were carried for 66, 79, and 33 data points, respectively. These data points were obtained by taking every 50[th] data point of molecules in the last 3 loops that had a ΔE or EA small than their mean minus standard deviation and an IP larger than their mean plus standard deviation of the model. In this way, 99, 86, 71 geometries were obtained for ΔE, EA, and IP, respectively. These were, as done in the original data set, optimized with PBE+vdW using first light and later tight settings of the basis set. The relaxed geometry was then used to compute G0W0@PBE0 values at the complete basis set limit. Therefore, two calculations were carried out, once with the QZVP basis set and once with the TZVP basis set. G0W0@PBE0 values at the complete basis set limit were extrapolated from TZVP and QZVP quasiparticle energies by a linear fit using the procedure employed for the GW100 benchmark set[11] with the script obtained from NOMAD of ref. [2]. Out of all calculations, 66, 79, and 33 converged for ΔE, EA, and IP, respectively. The reference values are plotted against the SchNet+H predictions in **Supplementary Figure 4**. In addition, the G0W0@PBE0 values of the original data set are shown. It is clearly visible that molecules predicted in the last iterations of the biasing process exhibit properties at the edges or outside of the training set.

As can be seen, SchNet+H accurately predicts the trends of almost all molecules correctly. There is one data point for the EA, which is predicted with a large error. The mean absolute error for ΔE, EA, and IP of molecules of the last biasing steps are 0.4 eV, 0.6 eV, and 0.4 eV, respectively. Given the fact that these molecules are at the edge of the originally learned distribution exhibiting electronic properties outside the training set and the use case of computationally efficient high-throughput screening, the accuracy can be deemed sufficient.

The smallest ΔE value computed with G0W0@PBE0 was 3.2 eV, while the smallest ΔE value of the OE62 data set is 4.8 eV, which is 1.6 eV larger. The mean ΔE value of the molecules recomputed is 3.9 eV, which is still smaller than the smallest value found in the OE62 data set. The mean ΔE value of the OE62 data set is 8.1 eV.

The largest EA value computed with G0W0@PBE0 was 6.6 eV, while the largest EA value of the OE62 data set is 2.4 eV, which is 4.2 eV larger. The mean EA of the molecules recomputed is 4.6 eV, which is still much larger than the largest value found in the OE62 data set. The mean EA of the OE62 data set is -0.7 eV.

The smallest IP computed with G0W0@PBE0 was 4.2 eV, while the smallest IP of the OE62 data set is 5.0 eV, which is 0.8 eV larger. The mean IP of the molecules recomputed is 5.0 eV, while the mean IP of the OE62 data set is 7.4 eV.

## Supplementary Section 4     Iterative biasing

For biasing of G-SchNet towards large EA, we selected all molecules with a target property, P, that was larger than the mean of each property, $\bar{P}$, plus the corresponding standard deviation, $\sigma_P$: $P = \bar{P} + x \cdot \sigma_P$ . For biasing of G-SchNet towards small IP, ΔE, and SCScore, we selected all molecules with a target property, P, that was larger than the mean of each property, $\bar{P}$, minus the corresponding standard deviation, $\sigma_P$: $P = \bar{P} - x \cdot \sigma_P$. For single property biasing we set x to 1. In case of biasing towards two properties, x was set to 0.5. The number of valid molecules generated in each loop and the number of molecules selected for biasing G-SchNet are shown in **Supplementary Datafile 1**.

## Supplementary Section 5     Computational costs of quantum chemistry calculations and machine learning training and predictions

The computational costs for G0W0@PBE0 and SchNet+H quasiparticle energies are compared in **Supplementary Table 1**. As can be seen, the computational costs of G0W0@PBE0 calculations are extremely large with several 1000 CPUhs for molecules larger than 80 atoms. The computational costs for SchNet+H predictions are almost independent of atom size and are averaged from predictions made for over 10k molecules. Dell PowerEdge C6420 compute nodes each with 2 x Intel Xeon Platinum 8268 (Cascade Lake) were used for molecules with up to about 45 atoms and Dell PowerEdge R640 nodes each with 2 x Intel Xeon Platinum 8268 (Cascade Lake) were used for larger molecules. SchNet+H predictions were carried out on Dell PowerEdge R740 nodes each with 3 x NVIDIA RTX 6000 24 GB RAM GPUs.

As can be seen in **Supplementary Table 1** the screening of several hundred thousand molecules is computationally extremely costly and can be regarded as infeasible, especially because high memory nodes are necessary for molecules larger than about 45 atoms. In contrast, SchNet+H is computationally efficient enough to predict several hundred thousand molecules within less than a day. Note that the costs of obtaining G0W0@PBE0 calculations are more expensive than PBE0 calculations, because two calculations are carried out: The first step is the prediction of orbital energies at PBE0 level of theory and the second step is the correction of these energy levels with a Δ-ML model for G0W0@PBE0. Since two slightly differently trained SchNet+H models were executed each time G-SchNet generated structures were screened, one loop took approximately 2 days on a GPU. G-SchNet training on OE62 data took approximately 1 week, while biasing took less than 1 day on a GPU.

*Supplementary Table 1: Computational costs of quantum chemical calculations and machine learning predictions. The computational costs of calculating PBE0 orbital energies and G0W0@PBE0 quasiparticle energies at the complete basis set (CBS) limit with density functional theory and SchNet+H are compared for two molecules of different sizes. Dell PowerEdge C6420 compute nodes each with 2 x Intel Xeon Platinum 8268 (Cascade Lake) were used for molecules with up to about 45 atoms and Dell PowerEdge R640 nodes each with 2 x Intel Xeon Platinum 8268 (Cascade Lake) were used for larger molecules. SchNet+H predictions were carried out on Dell PowerEdge R740 nodes each with 3 x NVIDIA RTX 6000 24 GB RAM GPUs.*

| Type of calculation | Molecule size | QC [CPUh] | SchNet+H [GPUh] |
|---|---|---|---|
| PBE0 | 42 | 7.1 | $4.4 \cdot 10^{-5}$ |
| G0W0@PBE0 CBS | 42 | 502 | $1.8 \cdot 10^{-4}$ |
| PBE0 | 85 | 47.3 | $4.4 \cdot 10^{-5}$ |
| G0W0@PBE0 CBS | 85 | 4,126 | $1.8 \cdot 10^{-4}$ |

## Supplementary Section 6     Clustering and principal component analysis (PCA)

The variance covered by the first 5 principal components using descriptors of molecules of the OE62 data and of all molecules as input are shown in **Supplementary Figure 6**.

In addition to the representation of the chemical space spanned by principal components obtained from the OE62 data set and the structural descriptors, we carried out PCA using bonding descriptors of the OE62 data set. The chemical space spanned by the OE62 data represented by the first two principal components of the bonding descriptors can be seen in **Supplementary Figure 5**. The plots verify results found by using structural descriptors (Figure 2b, d, and f) and suggest similar relevant regions in chemical space for small fundamental gaps and large electron affinities and different important regions in chemical space that make up small ionization potentials. Also here, we can see that generated molecules are within the regions covered by molecules in the OE62 data set.

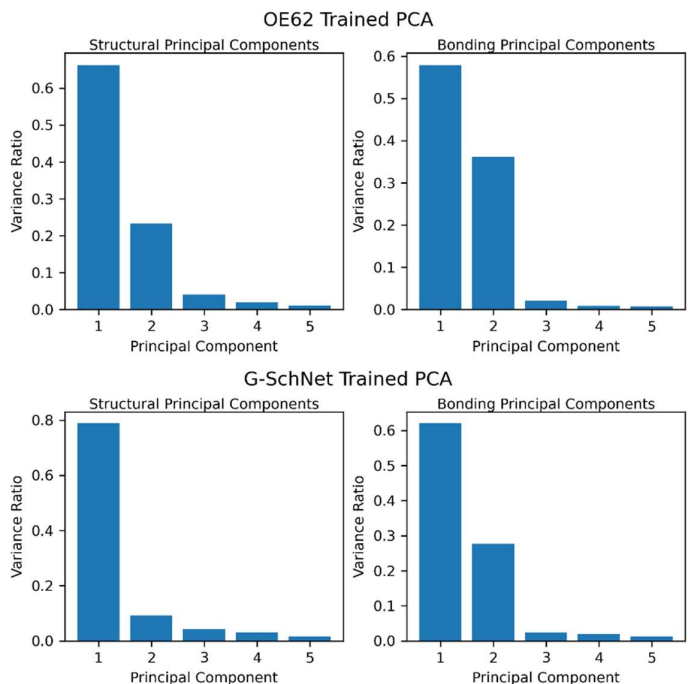

*Supplementary Figure 6: Variance in principal components. a) Variance of the first 5 principal components (PCs) obtained for the structural descriptor, i.e., SOAP, b) and the bonding descriptor, for molecules of the OE62 data set. c) Variance of the first 5 principal components (PCs) obtained for the structural descriptor, i.e., SOAP, d) and the bonding descriptor, for molecules of the OE62 data set and the generated molecules used for biasing towards small fundamental gaps.*

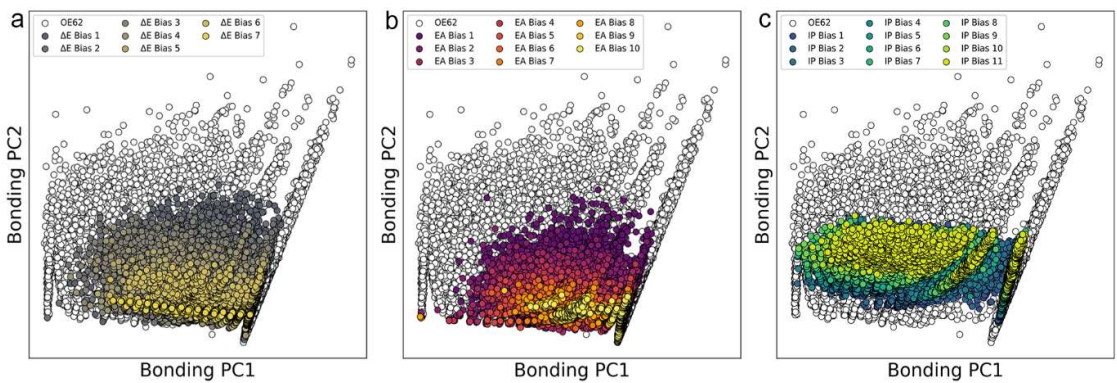

*Supplementary Figure 5: Chemical space spanned by OE62 data. Distribution of data points in chemical space made up by principal components obtained from OE62 data using bonding descriptors and results from biasing towards a) small fundamental gaps, ΔE, b) large electron affinities, EA, and c) small ionization potentials, IP. The color code indicates the biasing step. The plots are complementary to Figure 2 in the main text panels b, d, and f.*

**Supplementary Figure 7** shows the clusters plotted against the first principal components (PCs) obtained from structural descriptors and ΔE colored with respect to the loops (panel a) and clusters found (panel b). The subclusters are shown in panel c.

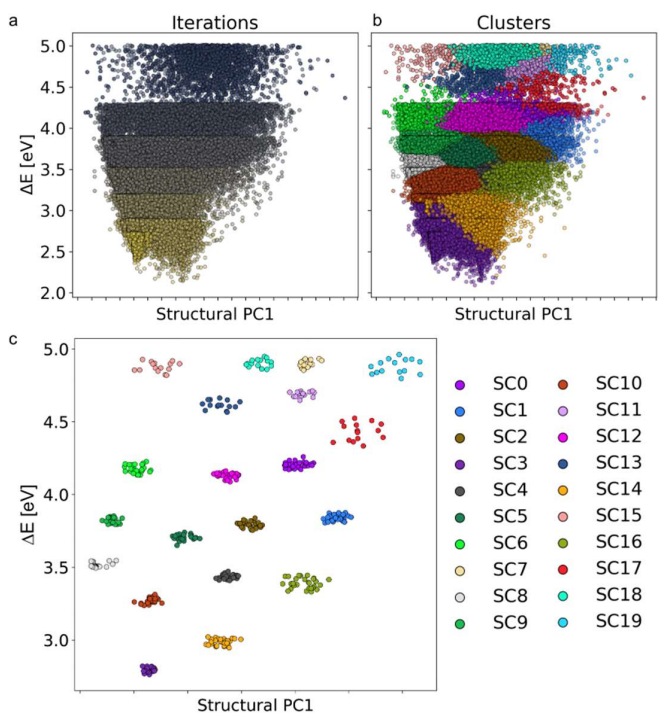

***Supplementary Figure 7: Clustering analysis for biasing G-SchNet towards small fundamental gaps, ΔE. A)** Data points obtained from OE62 and G-SchNet colored according to iterations and **b)** colored according to clusters found. **C)** 10 representatives of each cluster obtained with subclustering using centroids of b) as inputs SC indicates sub cluster.*

## Supplementary Section 7    Molecular features

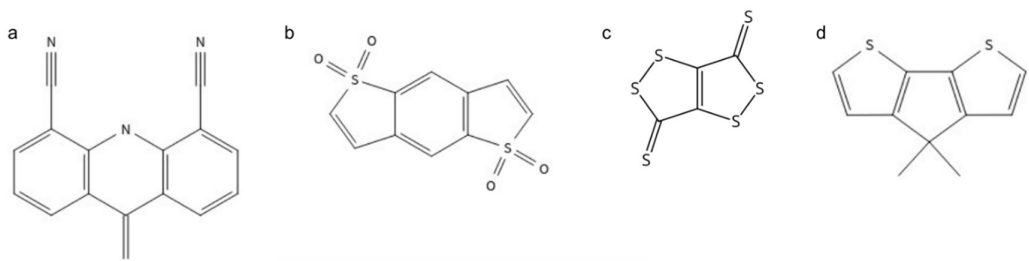

*Supplementary Figure 8: Functional groups represented in molecules with small fundamental gaps. Molecules strongly represented in the data set biased towards small fundamental gaps and generated with G-SchNet that are also found in the data set in Ref.44.*

**Supplementary Figure 8** shows functional groups that are represented frequently in molecules that have a small ΔE value. These molecular groups are parsed in SciFinder and are found in applications and are discussed in the main text.[12-14]

**Supplementary Section 8    Knock-out study**

To analyze whether G-SchNet can predict bonding patterns that are not present in the original data set, we eliminate all molecules containing cyano groups of the OE62 data set. These are molecules that have a C-N bond length of less than 1.25 Å, as C-N triple bonds are usually in the range of 1.15 Å. The modified OE62 data set is used to train a new G-SchNet model, which is then used to predict new molecules and is biased against small ΔE. As can be seen in **Supplementary Figure 9**a, the ΔE values iteratively decrease, when biased against them, which is expected. Supplementary Figure 9b shows that already after the first biasing step, G-SchNet predicts molecules with increased number of cyano groups. The trend of increased number of cyano groups in molecules with small ΔE values is thus retained.

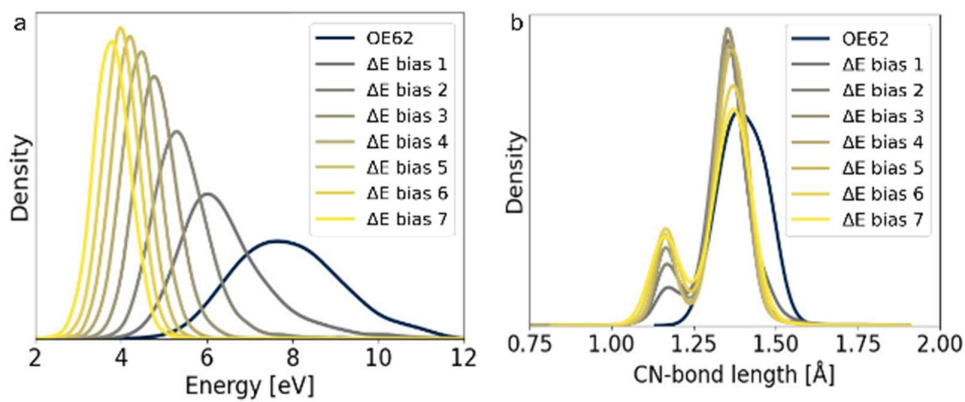

*Supplementary Figure 9: Knock-out study.* **A)** *Distribution of fundamental gaps, ΔE, and* **b)** *C-N bond lengths of molecules in the OE62 data set excluding molecules with a C-N bond length < 1.25 Å and of molecules generated with G-SchNet biased against ΔE.*

**Supplementary Section 9    Multi-property biasing**

As discussed in the main text in section 3.2 and 3.3, the synthetic complexity of molecules increases when minimizing the fundamental gap (see Figure 3h). This effect seems to revert after the third loop, when the complexity of synthesizability drops and becomes more favorable towards the end of the biasing process. However, it does not return to its original, lower distribution. This lowering of the complexity of synthesizability is possibly due to the fact that molecules become smaller with iterations, which generally reduces synthetic complexity.[15] The conclusion that our method is successful in finding rules in molecules that could be potentially relevant to optoelectronics, but that the molecules we generate are possibly too complex to synthesize, is not very encouraging. Therefore, we further sought to investigate the potential of the method to simultaneously optimize multiple properties, i.e., small fundamental gaps and low synthetic complexity of molecules.

The molecules selected for biasing G-SchNet initially are shown in **Supplementary Figure 10**. This image shows the fundamental gap against the SCScore of 340k molecules obtained from the OE62 data set and predicted with G-SchNet. The orange distribution is used for biasing G-SchNet initially.

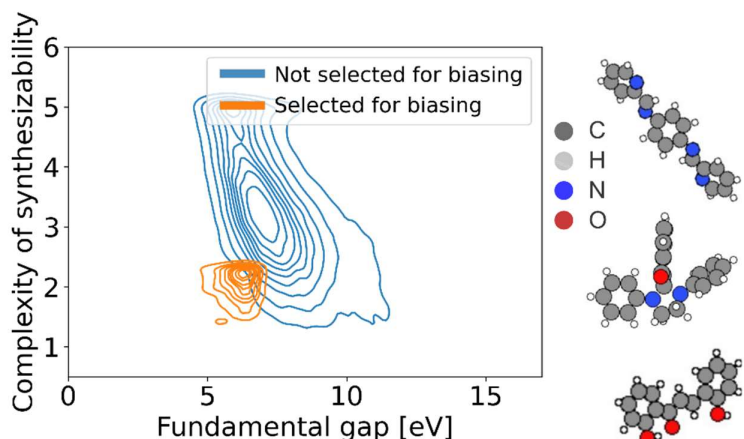

*Supplementary Figure 10: Molecules selected for multi-property biasing. Fundamental gap of molecules plotted against synthetic complexity score (SCScore) of molecules of the OE62 data set and generated with G-SchNet trained on the OE62 data set (blue distribution). The distribution of molecules selected for biasing towards small fundamental gaps are shown in orange. Some example molecules with small fundamental gaps and synthetic complexity (orange area) are shown right to the plot.*

The results, i.e., the sulfur nitrogen and selenium content (panel a), the elemental distribution in molecules (panel b), and the C-N bond lengths (panel c) are shown in **Supplementary Figure 11**. In addition to **Figure 4** in the main text. The plots are complementary to **Figure 4** in the main text, but contain results obtained by multi-property biasing, i.e., biasing towards small fundamental gaps and small SCScore, instead of results obtained only from biasing towards a single property, i.e., small fundamental gaps

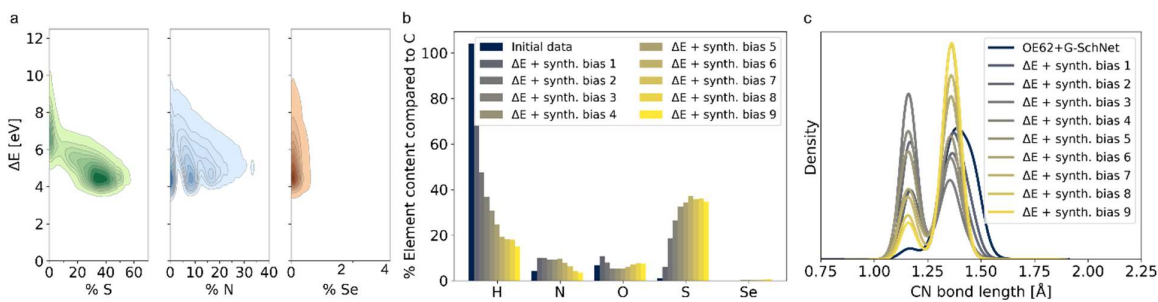

*Supplementary Figure 11: Cluster analysis for molecules with small fundamental gaps and small SCScore. a) Distribution of sulfur (S), nitrogen (N), and selenium (Se), b) elemental distribution and c) distribution of C-N bond lengths of molecules generated during biasing towards small fundamental gaps, ΔE, and small synthetic complexity score (SCScore).*